\documentclass[prl,preprint,showpacs,floatfix]{revtex4-1}
\usepackage{amsmath, amssymb, graphicx,subfigure}
\usepackage[latin1]{inputenc}
\usepackage{amsthm}
\usepackage[english]{babel}

\usepackage{verbatim}
\usepackage{ulem}
\usepackage{cancel}
\usepackage[colorlinks=true,linkcolor=blue,citecolor=blue]{hyperref}

\renewcommand\Re{\operatorname{Re}}
\renewcommand\Im{\operatorname{Im}}

\begin{document}
\title{Timing jitter in passively mode-locked semiconductor lasers}
\author{A. Pimenov$^{1}$\thanks{Corresponding author: pimenov@wias-berlin.de}, T. Habruseva$^{2,3,4}$,
D. Rachinskii$^{5,6}$, S.~P.~Hegarty$^{2,3}$, G. Huyet$^{2,3,7}$, and A. G. Vladimirov$^{1}$}
\affiliation{$^{1}$Weierstrass Institute, Mohrenstrasse 39, D-10117 Berlin, Germany}
\affiliation{$^{2}$Centre for Advanced Photonics and Process Analysis, Cork Institute of Technology,Cork, Ireland}
\affiliation{$^{3}$Tyndall National Institute, University College Cork, Lee Maltings, Dyke Parade, Cork, Ireland}
\affiliation{$^{4}$Aston University, Aston Triangle, B4 7ET, Birmingham, UK}
\affiliation{$^{5}$Department of Mathematical Sciences, The University of Texas at Dallas, US }
\affiliation{$^{6}$Department of Applied Mathematics, University College Cork, Ireland }
\affiliation{$^{7}$National Research University of Information Technologies, Mechanics and Optics, St Petersburg, Russia}

%%%%%%%%%%%%%%%%%%%%%%%%%%%%%%%%%%%%%%%%%%%%%%%%%%%%%%%%%%%%%%%%%%%%%%
%                                                                    %
%                         Platz fuer neue Theoreme                   %
%                                                                    %
%%%%%%%%%%%%%%%%%%%%%%%%%%%%%%%%%%%%%%%%%%%%%%%%%%%%%%%%%%%%%%%%%%%%%%
\begin{abstract}
We study the effect of noise on the dynamics of passively mode-locked semiconductor lasers both experimentally and theoretically. A method combining analytical and numerical approaches for estimation of pulse timing jitter is proposed. We investigate how the presence of dynamical features such as wavelength
bistability affects timing jitter.
%study how the nonlinear dynamical effects due to various physical characteristics of the laser devices can influence timing jitter.
%By varying the injection current we demonstrate the existence of bistability
%between two fundamental ML regimes.
%Moreover,
\end{abstract}


\maketitle
\section{Introduction}
Semiconductor mode-locked lasers received much attention in the last decade due to their multiple potential applications including high speed optical telecommunications and clocking~\cite{Delfyett_JLT06, Silva_JLT11}.
Pulses generated by these lasers are affected by noise due to spontaneous emission and other factors such as cavity optical length fluctuations. In particular,
their temporal positions in a pulse train deviate from those of the perfectly periodic output. This phenomenon called timing jitter
limits the performance of mode-locked devices \cite{Jiang2002}. Passive mode-locking is an attractive technique for periodic short pulse generation due to simplicity 
of implementation and handling as compared to other techniques such as hybrid or active mode-locking. 
However, in the absence of external reference clock passively mode-locked lasers exhibit relatively large pulse timing jitter \cite{Lin2010}. 
Bistable semiconductor lasers can be used as a basic element for optical switches \cite{pimenov2014,silverman2012} where the timing jitter can play a significant role as well.

Analytical approach to the study of
the influence of noise on mode-locked pulses propagating in the laser cavity was developed by H. Haus and A. Mecozzi \cite{haus1993}. Later this technique was extended to include the effects
of carrier density in semiconductor lasers \cite{haus2001}. However, many simplifications involved in the analysis of Refs. \cite{haus1993,haus2001} limit its applicability to modelling  dynamics of semiconductor lasers under the influence of noise.
In the last two decades extensive numerical simulations of travelling wave \cite{zhu,mork} and delay differential \cite{otto} models were performed to study timing jitter for different laser device configurations.
In particular, a monotonous decrease of pulse timing jitter with the increase of injection current in a passively mode-locked semiconductor laser was demonstrated both numerically and experimentally \cite{zhu}.
%The master equation has secant-shaped ML pulse as a steady state solution, and a small perturbation from this state can be studied using linearized equation of motion.
%The perturbed pulse is described by four parameters: the amplitude perturbation and the perturbations of phase, frequency, and timing. Using the orthogonality of the solutions of linearized equation
%of motion to the solutions of the adjoint system, linear coupled first order differential equations of motion, driven by noise, can be written out.
% Such approach, however, cannot be directly applied to the design of a particular device due to multiple simplifying assumptions,
%Purely computational approach is time costly, and the influence of noise on the dynamics of ML pulse has been studied in very limited parameter regions.
In this paper using a delay differential equation (DDE) model of a two-section passively mode-locked semiconductor laser \cite{VT04,VT05,vladimirov} 
we study the effect of noise on the characteristics of fundamental mode-locked regime and develop a semianalytical method for estimation of pulse timing jitter, which helps us to avoid high computational cost of a purely numerical approach.
With the help of DDE-BIFTOOL \cite{DDEbiftool} we perform numerical bifurcation analysis of the model and demonstrate the existence of
%for the first time the origin of the
bistability between two fundamental mode-locked regimes with different pulse repetition frequencies. By varying the injection current applied to
the gain section we demonstrate both experimentally and theoretically
%using the derived formula 3
three types of the injection current/timing jitter dependence: monotonous decrease of the pulse timing jitter with the increasing injection current,
peaks of timing jitter at certain intermediate values of injection current corresponding to bifurcation points, 
and abrupt drop of the timing jitter level after a transition between two stable branches of fundamental mode-locked regime occurs. We find both theoretically and experimentally, in some laser samples,
timing jitter can be reduced significantly
with the additional increase of the injection current, which also shifts the operation frequency.
%In Section \ref{sec:method} we consider an autonomous DDE model of a laser that operates in a ML periodic regime and a small perturbation from this state similarly to \cite{rebrova, arkhipov}. %$(A_0, G_0, Q_0)$
%We linearise the autonomous system near the ML periodic solution and project the perturbation term on the neutral eigenfunctions that correspond to the time shift and phase invariance
%of the autonomous equations in line with the Floquet theory of DDEs \cite{halanay, hale}.
% we compare the semianalytical asymptotic formula to the results of direct numerical simulations.
%In Section \ref{sec:results} using the derived formula in combination with DDE-BIFTOOL \cite{DDEbiftool} we study the dependence of timing jitter on various parameters of the system with a special attention to parameter regions near instabilities of the ML regime
%and compare our results with experiments.
%We confirm earlier observations that the increase of pumping into the gain section can decrease timing jitter, and demonstrate that with higher linewidth enhancement
%factor in the gain section the timing jitter can increase with the pumping. We investigate how instabilities of ML regime affect timing jitter.

\section{Method} \label{sec:method}
\subsection{Delay differential model}
We consider a delay differential equation (DDE) model of a two-section passively mode-locked semiconductor laser introduced in \cite{VT04,VT05,vladimirov}:
\begin{gather}
\partial _{t}A + \gamma A =   \gamma \sqrt{\kappa }%
\exp\biggl\{\frac{1-i\alpha_{g}}2 G\left( t-\tau\right)- \notag \\ \frac{ 1-i\alpha_{q}}2 Q\left( t-\tau\right) \biggr\}A\left(
t-\tau\right)+  \xi \eta(t),  \label{eq:DDE1} \\
\partial _{t}G = g_{0}-\gamma _{g}G
-\left( e^{G }-1\right) \left\vert A
\right\vert ^{2},  \label{eq:DDE2} \\
\partial _{t}Q = q_{0}-\gamma _{q}Q -s\left(
1-e^{-Q }\right) e^{G }\left\vert A \right\vert ^{2},  \label{eq:DDE3}
\end{gather}%
where $A(t)$ is the electric field envelope, $G(t)$ and $Q(t)$ are
 saturable gain and loss introduced by the gain and absorber sections correspondingly,
$\tau$ is the cold cavity round-trip time, $g_0$ and $q_0$ are unsaturated gain (pump) and absorption parameters, $\alpha_{g,q}$ are linewidth enhancement factors in the gain 
and absorber sections, $\gamma_{g,q}$ are carrier relaxation rates, $\gamma$ is the spectral filtering bandwidth, $s$ is a saturation parameter, $\kappa$ is the 
linear attenuation factor per cavity round trip, and $\eta(t)$ is $\delta$-correlated Langevin noise of amplitude $\xi$ \cite{otto}:
$$\eta(t) = \eta_1(t) + i \eta_2(t), \ \left\langle \eta_i(t)\right\rangle = 0,\ \left\langle\eta_i(t)\eta_j(t')\right\rangle=\delta_{i,j}\delta(t-t').$$
We denote by $\psi(t) = (\Re A, \Im A, G, Q)^T$ a real-valued solution of \eqref{eq:DDE1}. %The values of pararameters are given in the following table

\subsection{Estimation of timing jitter}
Due to time shift invariance of the autonomous system of equations \eqref{eq:DDE1}-\eqref{eq:DDE3} the timing phase of the solution $\psi(t)$ exhibits a random walk under the influence of the white noise \cite{mork, otto}. 
This leads to the appearance of the pulse timing jitter.
Hence, in order to estimate timing jitter one needs to separate the phase noise from the amplitude noise.
In numerical simulations pulse timing jitter can be estimated directly by calculating the variance of the temporal pulse positions $\sigma_{var}$
%of the field amplitude $|A(t)|$
selected in a specific manner from the time series of the electric field $A(t)$ \cite{haus2001,mork}.
%or using a technique called RMS timing jitter $\sigma_{RMS}$, which is mainly used by experimentalists and deals with the power spectrum of the pulse train
Experimentally, the timing jitter is usually estimated using the root mean square (RMS), $\sigma_{RMS}$, of the power spectrum of the pulse train
 \cite{zhu, mork, otto}.
Estimation of the variance of temporal pulse positions $\sigma_{var}$ can be done experimentally as well using a second-harmonic non-collinear optical cross-correlation technique, where the relation $\sigma_{RMS} \approx 2.3548 \sqrt{2} \sigma_{var}$ holds \cite{haus2001}.

\begin{figure}[h]
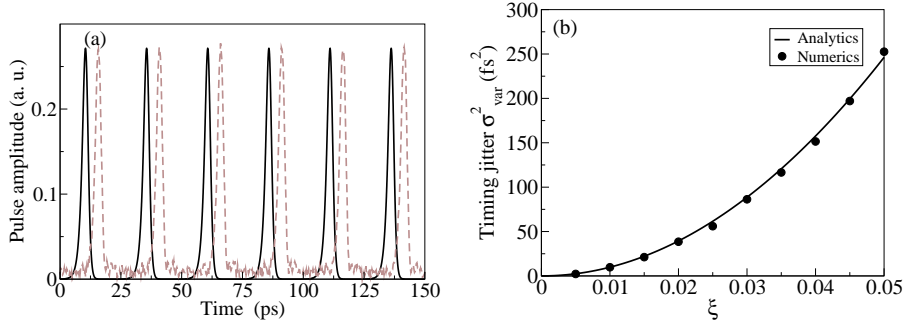

\centering{}%
\subfigure{ \includegraphics[clip,width=0.35\textwidth]{pulses_1}
\label{fig0a}}
\subfigure{ \includegraphics[clip,width=0.35\textwidth]{jitter}
\label{fig0b}}
\caption{(a) Time traces of the mode-locked solution of Eqs.~\eqref{eq:DDE1}-\eqref{eq:DDE3} at $\xi = 0$ (solid) and $\xi = 0.05$ (dashed).
(b) Comparison of results obtained using the asymptotic formula~\eqref{eq:jitter} (line) and those of numerical pulse timing jitter calculation from the amplitude time traces (circles) for different values of noise strength $\xi$.
Other parameters are $\tau=25$ ps, $\kappa=0.3$, $\gamma^{-1} =  125$ fs, $\gamma_g^{-1} = 500$ ps, $\gamma_q^{-1} = 5$ ps, $q_0^{-1} = 10$ ps, $g_0^{-1}=250$ ps, $\alpha_q = 1$, $\alpha_g=2$, $s = 50$.}
\label{fig0}
%In Figure~\ref{fig3} we can see that the timing jitter for the ML branch that exists for smaller $g_0$ is higher than for the other ML branch similarly to the experimental results shown in Fig.~\ref{fig:esa}.

\end{figure}

In this work we develop a semianalytical method for estimation of pulse timing jitter $\sigma_{var}$ in DDE model \eqref{eq:DDE1}-\eqref{eq:DDE3} using a perturbation technique similar to the method proposed by H. Haus and A. Mecozzi \cite{haus1993}.
We consider a locally stable fundamental mode-locked solution $\psi_0(t) = (\Re A_0, \Im A_0, G_0, Q_0)^T$ of system
\eqref{eq:DDE1}-\eqref{eq:DDE3} with $\xi = 0$ (see Fig.~\ref{fig0a}). This solution satisfies the conditions $|A_0(t)|^2=|A_0(t-T_0)|^2$, $G_0(t)=G_0(t-T_0)$, and $Q_0(t)=Q_0(t-T_0)$, 
where the mode-locked pulse period $T_0$ is close to the cavity round trip time $T$. Now let us suppose that $\xi \ll 1$.
The local dynamics of trajectories of
equations \eqref{eq:DDE1}-\eqref{eq:DDE3} near the periodic solution can be studied using linearisation of the system. 
Substituting $\psi(t) = \psi_0(t) + \delta \psi(t)$ into \eqref{eq:DDE1}-\eqref{eq:DDE3} and assuming that  $|\delta \psi| \ll 1$ we obtain the following linearized equation:
\begin{equation}\label{eq:linear}
-\frac{d}{dt} \delta \psi(t) + B(t) \delta \psi(t) + C(t - \tau) \delta \psi(t- \tau) + w(t) = 0,
\end{equation}
where $B$ and $C$ are the Jacobi matrices of the linearisation \cite{arkhipov}, $w(t) = (\xi \Re \eta(t),\xi \Im \eta(t),0,0)^T$.
The homogeneous system \eqref{eq:linear} with $w(t) \equiv 0$ has two linearly independent
periodic solutions,   the so-called neutral modes $\delta \psi_{0\theta} = d \psi_0(t)/dt$ and $\delta \psi_{0 \varphi} = (-\Im A_0, \Re A_0, 0, 0)^T$, 
which correspond to the time shift and phase shift symmetry of the system \eqref{eq:DDE1}-\eqref{eq:DDE3}, respectively.
Next, we project the increment $\psi(T_0) - \psi(0) = \delta \psi(T_0) - \delta \psi(0)$ of the perturbed solution $\psi(t)$ over the period on the neutral eigenfunction
$\delta \psi_{0\theta}$ \cite{haus1993} using the variation of parameters formula for DDEs \cite{hale}.
%We eliminate the impact of the amplitude noise due to the
%small stochastic perturbation $f(t)$ on the fluctuation of the period of ML solution by
%projecting the function $\delta \psi(T_0)$ on the neutral eigenfunction corresponding to time-shift invariance of the system \cite{haus1993} using the
%variaton of constants formula for DDE \cite{halanay, hale}.
%The standard analytical tools such as $L^2$-scalar product are, however,
%not directly applicable to the decomposition of the eigenspace of periodic solution of DDE models. Instead, we use the projection technique (variation of constants formula) proposed in the theory
%of DDEs \cite{halanay, hale}.

Let $\delta \psi$ be a solution of homogeneous problem \eqref{eq:linear} with $w(t) \equiv 0$ and a row vector
$\delta \psi^\dagger(t) = (\delta \psi^\dagger_{1}, \delta \psi^\dagger_{2},\delta \psi^\dagger_{3}, \delta \psi^\dagger_{4})$  be a solution of the adjoint problem
\begin{equation}\label{eq:linearadj}
\frac{d}{dt} \delta \psi^\dagger(t) + \delta \psi^\dagger(t) B(t) + \delta \psi^\dagger (t+ \tau) C(t) = 0.
\end{equation}
For solutions $\phi$ of \eqref{eq:linear} and $\phi^\dagger$ of \eqref{eq:linearadj} we consider the following bilinear form \cite{halanay, hale}
\begin{equation}\label{eq:form}
[ \phi^\dagger, \phi](t) = \phi^\dagger(t) \phi(t) + \int_{-\tau}^0 \phi^\dagger(t+\theta+\tau) C(t+\theta) \phi(t+\theta) d\theta.
\end{equation}
%If $\delta \psi_0$ is the solution of \eqref{eq:linearh} and $\delta \psi_0^\dagger$ is the solution  of \eqref{eq:linearadj} then for all $t$ the following relation holds
% $$\frac{d \langle \delta \psi^\dagger_0, \delta \psi_0\rangle_\theta(t)}{dt} \equiv 0.$$
%Neutral eigenfunctions are periodic solutions of \eqref{eq:linear} with $f\equiv 0$, and they correpond to the symmetries of the system \eqref{eq:DDE1}-\eqref{eq:DDE3}: time shift invariance $\delta \psi_{0\theta} = \frac{d \psi_0(t)}{dt}$, and
%phase invariance $\delta \psi_{0 \varphi} = (-\Im A_0, \Re A_0, 0, 0)^T$.
Adjoint periodic eigenfunctions $\delta \psi_{0\theta} ^{\dagger}$ and $\delta \psi_{0\varphi} ^{\dagger}$ (neutral modes of \eqref{eq:linearadj}) are $T_0$-periodic and satisfy biorthogonality conditions
%\footnote{Such biorthonormality condition for eigenfunctions of \eqref{eq:linearh} due to form \eqref{eq:form} is discussed in \cite{hale}.}
\begin{equation}\label{eq:orth}
\left[\delta \psi_{0j} ^{\dagger},  \delta \psi_{0k}  \right] = \delta_{jk},
\end{equation}
where $j, k = \{\theta, \varphi\}$.

Timing jitter $\sigma_{var}$ is given by the variance of the pulse repetition period fluctuations of the solution $\delta \psi(t)$ projected on the neutral eigenfunction 
$\delta \psi_{0\theta}$ corresponding to the time shift invariance of the model equations. Using variation of constants formula \cite{hale} we obtain
$$
\sigma^2_{var} = \mbox{var}\left(\int_0^{T_0} \delta \psi_{0\theta}^\dagger(s) w(s) ds\right).$$
For the Langevin term $w(t)$, this expression results in the stochastic Ito integral
\begin{equation*}
\begin{split}
\sigma^2_{var} = \xi^2 \mbox{var}\biggl(\int_0^{T_0} \delta \psi_{0\theta, 1}^\dagger(s)  d W_1(s)+\\ \int_0^{T_0} \delta \psi_{0\theta, 2}^\dagger(s) dW_2(s)\biggr),
\end{split}
\end{equation*}
where $W_1$ and $W_2$ are two independent Wiener processes.
Since the expected value of this integrals is 0, finally, we obtain
\begin{equation}\label{eq:jitter}
\sigma^2_{var} = \xi^2 \int_0^{T_0} \left(\left( \delta \psi_{0\theta, 1}^\dagger(s)\right)^2  + \left(\delta \psi_{0\theta, 2}^\dagger(s)\right)^2 \right)ds.
\end{equation}
Fig. \ref{fig0b} indicates that the timing jitter values estimated with the help of asymptotic formula~\eqref{eq:jitter} are in good agreement with those obtained numerically from the field amplitude time traces \cite{mork, otto}.

\section{Experiment}
The experiments were performed with two-section monolithic quantum-dot mode-locked devices. The active region consisted of $15$ layers of InAs quantum-dots grown on GaAs substrate at Innolume GmbH. 
The devices were cleaved with no coatings applied to the facets and mounted on a temperature controlled
stage. We used a set of lasers with different absorber section lengths in the experiments.

Pulse timing jitter fluctuations were
measured via the integration of the normalized power spectral density (PSD), $L_{RF}(f)$, over a certain range~\cite{Linde_APB86}:
\begin{eqnarray}
\sigma^{i}(f_1,f_2)=\frac
{T_r}{2\pi}\sqrt{2\int_{f_1}^{f_2}L_{RF}(f)df},
\end{eqnarray}
where $\sigma^{i}$ is an integrated timing jitter, $f_1$ and $f_2$ are
the lower and upper integration limits, and $T_r$ is the pulse train period.

The single sideband PSD was measured from the output radio-frequency (RF) spectrum using an Advantest electronic spectrum analyzer (ESA), a fast photodetector, and an amplifier.
The normalized PSD is given by the expression:
\begin{eqnarray}
L_{RF}(f)= \frac{S_{RF}(f)}{RBW \times S_t},
\end{eqnarray}
where $S_t$ is the peak signal power, RBW is the resolution bandwidth of the ESA, and $S_{RF}(f)$ is PSD.

\begin{center}
\begin{figure}[h]
  \centering
  \includegraphics[width=0.65\textwidth]{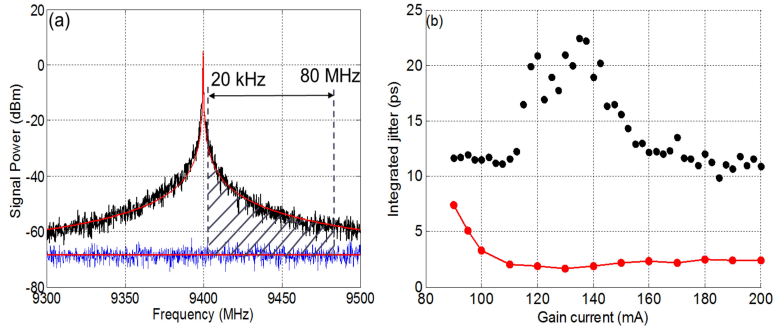}
  \caption{(a) Integration of timing jitter. Measured RF signal (black) and Lorentzian fit (red), laser $P_2$. (b) Integrated timing jitter for lasers $P_1$ (red) and $P_2$ (black) versus gain current.}
  \label{fig:TJ}
\end{figure}
\end{center}

The single sideband PSD was integrated over the range of $20$
kHz-$80$ MHz as shown in Fig.~\ref{fig:TJ}(a), from both sides. When the RF spectra fit well with the Lorentzian (Fig.~\ref{fig:TJ}(a), red) the PSD can be estimated from the Lorentzian fit and multiplied by two.

For the Lorentzian line shape, the integrated timing
jitter can be determined from the RF linewidth, $\Delta \nu_{RF,1}$, of the first harmonic in the RF spectrum and pulse train period $T_r$
~\cite{Kefelian_PTL08}:
\begin{eqnarray}\label{eqn:tjlor}
\sigma^{i}(f_1,f_2)= \frac {T_r\sqrt{\Delta
\nu_{RF,1}}}{2\pi^{3/2}}\sqrt{\frac{1}{f_1}-\frac{1}{f_2}}.
\end{eqnarray}

\begin{center}
\begin{figure}[h]
  \centering
  \includegraphics[width=0.6\textwidth]{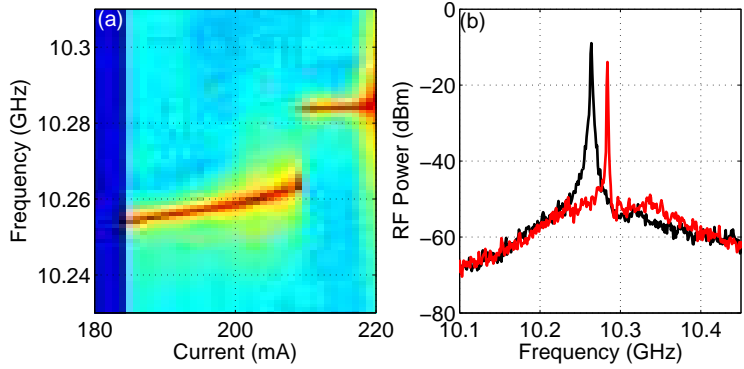}
  \caption{(a) Evolution of the RF spectrum with gain current. (b) RF spectra for the transition point for the gain current of $209$ mA (black) and $210$ mA (red). Laser $P_3$, reverse bias $-4.0$ V. }
  \label{fig:esa}
\end{figure}
\end{center}

Fig.~\ref{fig:TJ}(b) shows measured timing jitter for the lasers $P_1$ (red) and $P_2$ (black) with 17\% and 12\% absorber sections, respectively,
as a function of gain current. For the laser $P_1$ the jitter demonstrated monotonic decrease with the gain current, while for the laser $P_2$ the integrated jitter behavior was non-monotonic with values ranging from $11$ to $23$ ps.
%\onecolumngrid
%\twocolumngrid
Fig.~\ref{fig:esa}(a) shows measured RF spectrum evolution versus gain current. The transition between two mode-locked regimes can be seen at the gain current of around $209$ mA with a shift of the repetition rate and decrease of jitter. 
Examples of the RF spectra at the transition point are shown in Fig.~\ref{fig:esa}(b) by black and red lines for the gain currents of $209$ mA and $210$ mA, respectively. It can be seen that the RF linewidth after the transition point is about twice narrower.

%\begin{figure}[h!]
%  \centering
%  \includegraphics[width=0.65\textwidth]{F3v40_fig2.eps}
%  \caption{(a) Evolution of the RF spectrum with gain current.(b) RF spectra for two mode-locked regimes at the transition point for the gain current of $209$ mA (black) and $210$ mA (red). Lasers $P_3$ with ... absorber section. Reverse bias $-4.0$ V.}
%  \label{fig:esa}
%\end{figure}
%\onecolumngrid
%\twocolumngrid
%\begin{figure}[h]
%\centering{}%
%\subfigure{ \includegraphics[clip,width=0.47\textwidth]{TJmeas_2a}
%\label{fig:TJmeasure}}
%\subfigure{ \includegraphics[clip,width=0.43\textwidth]{TJ_LlLj}
%\label{fig1b}}
%\caption{(a) Integration of timing jitter. RF signal (black) and Lorentzian fit (red), laser $P_2$. (b) Integrated timing jitter for lasers $P_1$ (red) and $P_2$ (black) versus gain current.}
%\label{fig1}
%\end{figure}

\section{Numerical results} \label{sec:results}
%The model \eqref{eq:DDE1}-\eqref{eq:DDE3} is too simple to represent accurately the variety of dynamics of quantum dot lasers,
%however it is possible to get some qualitative insights into their complex behaviour by introducing asymmetry in the phase-amplitude coupling in the gain and absorber
%sections using different linewidth enhancement factors $\alpha_g > \alpha_q$ [pimenov2014].
For our numerical analysis we chose the parameters of model equations \eqref{eq:DDE1}-\eqref{eq:DDE3} to match those of a 10 GHz laser \cite{pimenov2014} :
$\tau=100 ps$, $\gamma^{-1} =  0.5$ ps, $\gamma_g^{-1} = 500$ ps, $\gamma_q^{-1} = 10$ ps, $q_0^{-1} = 5.56$ ps, $\alpha_q = 1$, $\kappa = 0.3$, and the level of noise $\xi = 0.05$ \cite{otto}.

\begin{figure}[h]
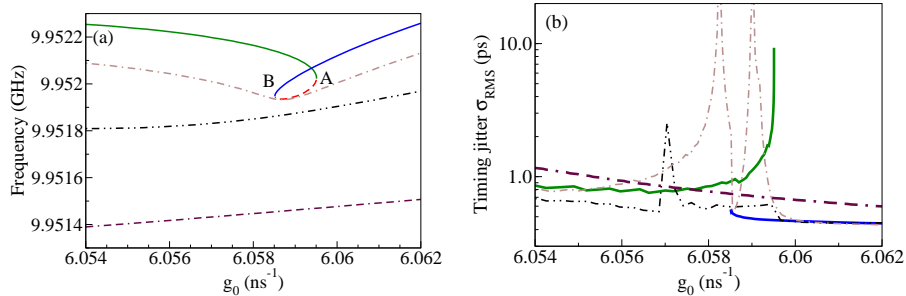

\centering{}%
\subfigure{ \includegraphics[clip,width=0.35\textwidth]{biftree_bistab_3_2}
\label{fig3a}}
\subfigure{ \includegraphics[clip,width=0.35\textwidth]{jitter_bistab_3_2}
\label{fig3b}}
\caption{Branches of fundamental ML regime (a) and corresponding timing jitter (b) for $s=24$ (dash-dash-dotted), $s = 11$ (dash-dot-dotted), $s=9.5$ (dash-dotted), $s = 8.9$ (solid and dashed for unstable), where A and B are fold bifurcation points.
Other parameters are $\gamma^{-1} =  0.5$ ps, $\gamma_g^{-1} = 500$ ps, $\gamma_q^{-1} = 10$ ps, $q_0^{-1} = 5.56$ ps, $\alpha_q = 1$, $\alpha_g=5$.}
\label{fig3}
%In Figure~\ref{fig3} we can see that the timing jitter for the ML branch that exists for smaller $g_0$ is higher than for the other ML branch similarly to the experimental results shown in Fig.~\ref{fig:esa}.

\end{figure}

It has been previously noted that high asymmetry in the phase-amplitude coupling in the gain and absorber sections can be strongly linked to complex dynamical behaviour of
semiconductor lasers \cite{vladimirov}. This strong coupling can be accounted for by assuming that the difference between linewidth enhancement factors $\alpha_g - \alpha_q$ in the gain and absorber sections is sufficiently large \cite{pimenov2014}.
We chose the pumping parameter $g_0$ %(see Fig.~\ref{fig2})
as the continuation parameter to obtain bifurcation diagrams in Fig.~\ref{fig3} with the help of
the DDE-BIFTOOL package \cite{DDEbiftool}. The timing jitter was estimated using Eq.~\eqref{eq:jitter} by implementing a MATLAB code for calculation of the eigenfunctions of the adjoint eigenproblem \eqref{eq:linearadj}.

Motivated by the timing jitter comparison performed in the experiment (see Fig.~\ref{fig:esa}) for the bistable branches of fundamental mode-locked regimes with different pulse repetition frequencies,
we find that a bistability between two mode-locked regimes exists in the model \eqref{eq:DDE1}-\eqref{eq:DDE3} for $g_0$ between $6.058$ and $6.06$ ns$^{-1}$, see Fig.~\ref{fig3}(a). 
We see from Fig.~\ref{fig3} (b) that for each of the two branches the timing jitter first decreases with increasing $g_0$ and then increases towards the instability threshold of the mode-locking regime. 
Similarly to the experimental results of Fig.~\ref{fig:esa}, timing jitter drops abruptly when the solution switches from the branch staring at the point A to the branch starting at the point B. 
When the saturation parameter $s$ is increased to 9.5 the fold bifurcations disappear (see dash-dotted line in Fig.~\ref{fig3}(a)) and a single branch of mode-locking regime remains stable. 
However, as it is illustrated by the dash-dotted line in Fig.~\ref{fig3}(b) the pulse timing jitter dependence on the pump parameter remains similar to that in the bistable case, 
demonstrating large peaks near the former instability points.  At even higher value of the saturation parameter $s=11$ the timing jitter dependence still has a smaller peak in the same region.
Our simulations performed with the help of the software package DDE-BIFTOOL indicate that these peaks appear when one of the negative Lyapunov exponents of the mode-locked solution $\psi_0(t)$ comes very close to zero
 with the change of the parameter $g_0$.
Such local increase in the pulse timing jitter obtained numerically for $\alpha_g=5$ is in agreement with the experimental results shown in Fig.~\ref{fig:TJ}(b) for the laser $P_2$. 
Finally, for $s=24$ we observed monotonous decrease of the pulse timing jitter with the increase of $g_0$, which is in agreement with the experimental data obtained for the laser $P_1$.

%\begin{figure}[h]
%\centering{}%
%\subfigure{ \includegraphics[clip,width=0.35\textwidth]{biftree_alp}
%\label{fig2a}}
%\subfigure{ \includegraphics[clip,width=0.35\textwidth]{jitter_alp}
%\label{fig2b}}
%\caption{
%Branches of fundamental ML regime (a) and corresponding timing jitter (b) for $\alpha_g=3.3, s = 10$ (thick dash-dotted), $\alpha_g=5, s=9.5$ (thin dashed), $\alpha_g = 5, s = 8.9$ (solid). Other parameters are
%$\gamma^{-1} =  0.5$ ps, $\gamma_g^{-1} = 500$ ps, $\gamma_q^{-1} = 10$ ps, $q_0^{-1} = 5.56$ ps, $\alpha_q = 1$.
%}
%\label{fig2}
%\end{figure}

\section{Conclusion} \label{sec:conclusion}
We have studied theoretically and experimentally the effect of noise on the characteristics of a two-section semiconductor laser operating in fundamental mode-locking regime. We have proposed a semianalytical method for estimation of timing jitter 
in a system of DDEs \eqref{eq:DDE1}-\eqref{eq:DDE3} describing this laser. The proposed method
can be applied to study noise characteristics of other multimode laser devices modeled by systems of  delay-differential equations \cite{otto, rebrova, pimenov2013}.
Using the software package DDE-BIFTOOL we have studied the dependence of the timing jitter on the injection current and other laser parameters and, in contrast to previous theoretical results \cite{zhu}, we have 
demonstrated that this dependence can be non-monotonous.  Specifically, we have observed a peak 
in the gain dependence curve of timing jitter, which is related to the presence of a real Lyapunov exponent approaching zero from below.
Our results suggest that in the bistable regime of operation the branch of mode-locking regime that is stable at lower injection currents exhibits higher level of pulse timing jitter. A sudden drop in the timing jitter is observed when the laser switches to another branch of mode-locking regime with the increase of the injection current. 

A. P. and A. G. V. acknowledge the support of SFB 787 of the DFG, project B5.
T.H. acknowledges support of Marie Curie Action FP7-PEOPLE-2011-IEF, HARMOFIRE project, Grant No 299288.
G. H., S. P. H. and A. G. V. acknowledge the support of EU FP7 Marie Curie Action FP7-PEOPLE-2010-ITN through the PROPHET project, Grant No. 264687.
A. G. V. and G. H. acknowledge the support of E. T. S. Walton Visitors Award of the
SFI.
G. H. and S. P. H. were also supported by the Science Foundation Ireland (SFI) under Contract No. 11/PI/1152, and under the framework of the INSPIRE Structured PhD Programme, funded by the Irish Government's Programme for Research in Third Level Institutions, Cycle 5, National Development Plan 2007-2013.

\bibliographystyle{plain}   %>>>> makes bibtex use spiebib.bst
\bibliography{noise}   %>>>> bibliography data in report.bib

\begin{thebibliography}{10}

\bibitem{arkhipov}
R.~Arkhipov, A.~Pimenov, M.~Radziunas, D.~Rachinskii, A.~G. Vladimirov,
  D.~Bimberg, and et~al.
\newblock Hybrid mode-locking in semiconductor lasers: simulations, analysis
  and experiments.
\newblock {\em IEEE Jornal of Selected Topics in QE}, 99:1100208, 2013.

\bibitem{Delfyett_JLT06}
P.~J. Delfyett, S.~Gee, M.-T. Choi, H.~Izadpanah, W.~Lee, S.~Ozharar,
  F.~Quinlan, and T.~Yilmaz.
\newblock Optical frequency combs from semiconductor lasers and applications.
\newblock {\em J. Lightwave Technol}, 24:27012719, 2006.

\bibitem{Silva_JLT11}
Marcia~Costa e~Silva, Alexandra Lagrost, Laurent Bramerie, Mathilde Gay, Pascal
  Besnard, Michel Joindot, Jean-Claude Simon, Alexandre Shen, and Guan-Hua
  Duan.
\newblock {Up to $427$ GHz All Optical Frequency Down-Conversion Clock Recovery
  Based on Quantum-Dash Fabry-Perot Mode-Locked Laser}.
\newblock {\em J. Lightwave Technol.}, 29 (4), 2011.

\bibitem{DDEbiftool}
K.~Engelborghs, T.~Luzyanina, and G.~Samaey.
\newblock {DDE-BIFTOOL} v.2.00: A {MATLAB} package for bifurcation analysis of
  delay differential equations.
\newblock Technical Report TW-330, Department of Computer Science, K.U.Leuven,
  Leuven, Belgium, 2001.

\bibitem{halanay}
A.~Halanay.
\newblock {\em Differential Equations: Stability, Oscillations, Time Lags}.
\newblock Academic Press, 1966.

\bibitem{hale}
J.~Hale.
\newblock {\em Theory of Functional Differential Equations}.
\newblock Springer-Verlag, 1977.

\bibitem{haus1993}
H.~A. Haus and A.~Mecozzi.
\newblock Noise of mode-locked lasers.
\newblock {\em IEEE JQE}, 29(3):983--995, 1993.

\bibitem{haus2001}
L.~Jiang, M.~E. Grein, H.~A. Haus, and E.~P. Ippen.
\newblock Noise of mode-locked semiconductor lasers.
\newblock {\em IEEE Jornal of Selected Topics in QE}, 7(2):159--167, 2001.

\bibitem{Jiang2002}
Leaf~A. Jiang, Matthew~E. Grein, Erich~P. Ippen, Cameron McNeilage, Jesse
  Searls, and Hiroyuki Yokoyama.
\newblock Quantum-limited noise performance of a mode-locked laser diode.
\newblock {\em Opt. Lett.}, 27(1):49--51, Jan 2002.

\bibitem{Kefelian_PTL08}
F.~K{\'e}f{\'e}lian, S.~O'Donoghue, M.~T. Todaro, J.~G. McInerney, and
  G.~Huyet.
\newblock {RF Linewidth in Monolithic Passively Mode-Locked Semiconductor
  Laser}.
\newblock {\em IEEE Photon. Tech. Lett.}, 20 (16), 2008.

\bibitem{Lin2010}
Chang-Yi Lin, Frederic Grillot, Yan Li, Ravi Raghunathan, and Luke~F. Lester.
\newblock Characterization of timing jitter in a 5 ghz quantum dot passively
  mode-locked laser.
\newblock {\em Opt. Express}, 18(21):21932--21937, Oct 2010.

\bibitem{mork}
J.~Mulet and J.~Mork.
\newblock Analysis of timing jitter in external-cavity mode-locked
  semiconductor lasers.
\newblock {\em IEEE JQE}, 42(3):249--256, 2006.

\bibitem{otto}
C.~Otto, K.~L\"udge, A.~G. Vladimirov, M.~Wolfrum, and E.~Sch\"oll.
\newblock Delay-induced dynamics and jitter reduction of passively mode-locked
  semiconductor lasers subject to optical injection.
\newblock {\em New Journal of Physics}, 14:113033, 2012.

\bibitem{pimenov2013}
A.~Pimenov, V.~Z. Tronciu, U.~Bandelow, and A.~G. Vladimirov.
\newblock Dynamical regimes of a multistripe laser array with external off-axis
  feedback.
\newblock {\em J. Opt. Soc. Am. B}, 30(6):1606--1613, Jun 2013.

\bibitem{pimenov2014}
A.~Pimenov, E.~A. Viktorov, S.~P. Hegarty, T.~Habruseva, G.~Huyet,
  D.~Rachinskii, and A.~G. Vladimirov.
\newblock Bistability and hysteresis in an optically injected two-section
  semiconductor laser.
\newblock {\em Phys. Rev. E}, 89:052903, May 2014.

\bibitem{rebrova}
N.~Rebrova, G.~Huyet, D.~Rachinskii, and A.~G. Vladimirov.
\newblock Optically injected mode-locked laser.
\newblock {\em Physical Review E}, 83:066202, 2011.

\bibitem{silverman2012}
K.~Silverman, M.~Feng, R.~Mirin, , and S.~Cundiff.
\newblock Exotic behavior in quantum dot mode-locked lasers: Dark pulses and
  bistability.
\newblock In {\em Quantum Dot Devices}, volume~13 of {\em Lecture Notes in
  Nanoscale Science and Technology}, pages 23--48. Springer-Verlag, NY, 2012.

\bibitem{VT04}
A.~G. Vladimirov and D.~Turaev.
\newblock New model for mode-locking in semiconductor lasers.
\newblock {\em Radiophys. \& Quant. Electron.}, 47(10-11):857--865, 2004.

\bibitem{VT05}
A.~G. Vladimirov and D.~Turaev.
\newblock Model for passive mode-locking in semiconductor lasers.
\newblock {\em Phys. Rev. A}, 72:033808 (13 pages), 2005.

\bibitem{vladimirov}
A.~G. Vladimirov, D.~Turaev, and G.~Kozyreff.
\newblock Delay differential equations for mode-locked semiconductor lasers.
\newblock {\em Opt. Lett.}, 29:1221--1223, 2004.

\bibitem{Linde_APB86}
D.~{von der Linde}.
\newblock {Characterization of the Noise in Continuously Operating Mode-Locked
  Lasers}.
\newblock {\em Appl. Phys. B}, 39, 1988.

\bibitem{zhu}
B.~Zhu, I.~H. White, R.~V. Penty, A.~Wonfor, E.~Lach, and H.~D. Summers.
\newblock Theoretical analysis of timing jitter in monolithic multisection
  mode-locked dbr laser diodes.
\newblock {\em IEEE JQE}, 33(7):1216--1220, 1997.

\end{thebibliography}

\end{document}